\documentclass[twocolumn,amsfonts,showpacs,superscriptaddress,amsmath,amssymb,aps, longbibliography,prd]{revtex4-1}

\usepackage[makeroom]{cancel}
\usepackage{latexsym}
\usepackage{longtable}
\usepackage{epsfig}
\usepackage{comment}
\usepackage{graphicx,bbm,psfrag}
\usepackage{amssymb,amsmath,amsthm,booktabs,mathtools}
\usepackage{bm}
\usepackage{color}
\usepackage{dutchcal}
\usepackage{hyperref}
\usepackage{tikz}

\hypersetup{
    colorlinks,
    citecolor=red,
    linkcolor=blue,
}

\newcommand{\bc}{\begin{center}}
\newcommand{\ec}{\end{center}}
\def\ba#1{\begin{array}{#1}\displaystyle}
\newcommand{\ea}{\end{array}}

\newcommand{\beq}{\begin{equation}}
\newcommand{\eeq}{\end{equation}}
\newcommand{\beqa}{\begin{eqnarray}}
\newcommand{\eeqa}{\end{eqnarray}}

\newcommand{\n}{\nonumber\\}
\newcommand{\bi}{\begin{itemize}}
\newcommand{\ei}{\end{itemize}}

\def\b#1{\bar{#1}}
\def\frc#1#2{\frac{#1}{#2}}

\newcommand{\p}{\partial}

\newcommand{\R}{{\mathbb{R}}}

\newcommand{\ttbar}{$T\bar{T}$}

\newcommand{\varep}{\varepsilon}

\newcommand{\ri}{{\rm i}}

\newcommand{\dd}{{\rm d}}

\DeclareMathOperator{\sgn}{sgn}

\begin{document}

\title{New classical integrable systems from generalized $T\Bar{T}$-deformations}
 \author{Benjamin Doyon}
\affiliation
{Department of Mathematics, King’s College London, Strand, London WC2R 2LS, U.K.}
\author{Friedrich H\"ubner}
\affiliation{Department of Mathematics, King’s College London, Strand, London WC2R 2LS, U.K.}
\author{Takato Yoshimura}
\affiliation{All Souls College, Oxford OX1 4AL, U.K.}
\affiliation{Rudolf Peierls Centre for Theoretical Physics, University of Oxford,
1 Keble Road, Oxford OX1 3NP, U.K.}
%
%
\begin{abstract}
We introduce and study a novel class of classical integrable many-body systems obtained by generalized $T\Bar{T}$-deformations of free particles. Deformation terms are bilinears in densities and currents for the continuum of charges counting asymptotic particles of different momenta. In these models, which we dub ``semiclassical Bethe systems'' for their link with the dynamics of Bethe ansatz wave packets, many-body scattering processes are factorised, and two-body scattering shifts can be set to an almost arbitrary function of momenta. The dynamics is local but inherently different from that of known classical integrable systems. At short scales, the geometry of the deformation is dynamically resolved: either particles are slowed down (more space available), or accelerated via a novel classical particle-pair creation/annihilation process (less space available). The thermodynamics both at finite and infinite volumes is described by the equations of (or akin to) the thermodynamic Bethe ansatz, and at large scales generalized hydrodynamics emerge.
\end{abstract}

\maketitle

{\bf Introduction.}--- Since the inception of generalized hydrodynamics (GHD) in 2016 \cite{PhysRevX.6.041065,PhysRevLett.117.207201}, there has been a resurgence of interests in understanding the nature of the dynamics in integrable systems. GHD has proven to be a powerful and universal tool that captures it at large scales, and its predictions have been tested against cold-atom experiments in different platforms \cite{PhysRevLett.122.090601,doi:10.1126/science.abf0147,doi:10.1126/science.abk2397}. GHD has also been studied beyond the quantum realm and applied to classical systems including various types of hard rods \cite{Spohn1991,Doyon_2017,ferrari2022hard}, the Toda chain \cite{spohn_generalized_2019,doyon_generalised_2019}, the nonlinear Schrodinger equation \cite{10.1063/5.0075670,koch2022generalized}, the Calogero-Moser model \cite{bulchandani2021quasiparticle} and the sinh- and sine-Gordon models \cite{bastianello2018generalized,10.21468/SciPostPhys.15.4.140,bastianello2023sinegordon}. It provides the statistical framework \cite{Bonnemain_2022} for the theory of soliton gases \cite{El_2021,suret2023soliton}. The structure of GHD is extremely general, and it requires only limited data from the underlying system, such as the two-body scattering shift.

Yet, a full understanding of how GHD emerges from the microscopic dynamics is still lacking. In the hard-rods and  box-ball systems, rigorous proofs from slowly varying ensembles are available \cite{Boldrighini1983,2020cs,ferrari2022hard}; in soliton gases they are obtained from finite-gap solutions \cite{el2003thermodynamic,el2005,el2020spectral,El_2021}; ab initio derivations exist from kinetic theory \cite{PhysRevLett.128.190401} and Bethe-ansatz semiclassical principles \cite{doyon2023ab}; and the equations of state are well understood \cite{cubero2021form,borsi2021current}. But every model has specific properties; to understand the universality of GHD, it is important to construct {\em new integrable systems that can access the full space of GHD equations}.

In this letter, we do just that. We define a new class of classical many-particle systems with short-range interactions that are integrable, and that cover a {\em very large space of scattering functions}. The new systems are shown to arise from {\em generalised \ttbar-deformations}. \ttbar-deformations were introduced in relativistic quantum field theory \cite{Zamolodchikov:2004ce,Cavagli2016,smirnov_space_2017} as integrability-preserving deformations based on local conserved currents, that modify the scattering matrix by ``CDD factors" \cite{Cavagli2016,smirnov_space_2017}. Matrix elements of local fields have been recently studied \cite{castroalvaredo2023form,castroalvaredo2023form2,hotrepresentation2}, and \ttbar-deformations have been adapted to systems of different kinds \cite{Pozsgay2020,PhysRevE.104.044106,PhysRevE.104.064124,esper2021tt,cardy2022t,jiang2022mathrm,PhysRevLett.131.037101}, see the review \cite{jiang2021pedagogical}. Here ``generalised \ttbar-deformations" are those proposed in \cite{10.21468/SciPostPhys.13.3.072}, based on the larger space of {\em extensive} conserved charges first studied in the context of the non-equilibrium dynamics of integrable systems \cite{Ilievski_2016,Doyon2017,de2022correlation}. One admits conserved quantities measuring the density of asymptotic momenta,
and generalised \ttbar-deformations modify the scattering matrix by an {\em arbitrary momentum function}, although no explicit construction was made. Our models provide the explicit construction for classical Galilean particle systems. We confirm that they are Liouville integrable, that many-particle scattering is elastic and factorises into two-body shifts, and that the two-body shift can be chosen as an almost arbitrary function of momenta.

Constructing effective, fully integrable many-particle Hamiltonians for other objects such as solitons in nonlinear media is an old problem, see e.g.~\cite{PhysRevA.46.R2973}. Our models are the first to do that, and give in particular the potential for precise initial state preparation in inhomogeneous soliton gases \cite{El_2021,suret2023soliton}, a problem of current relevance.

In \cite{cardy2022t} it was shown that ``mass-momentum" \ttbar-deformations simply change the length of particles. This can also be interpreted as a local change of the {\em effective space} particles freely travel through, a special case of a geometric interpretation \cite{Conti2019} much like in GHD \cite{Doyon2018},
but the exact local properties of (generalised) \ttbar-deformation are nebulous.
We obtain an explicit microscopic dynamics implementing the geometry, generalising the change of particle lengths. Particles are tracers for their asymptotic momenta and ``go through" each other: additional available space is implemented by a slowing down at particles' proximity, while reduced space, by a novel process of creation and annihilation of pairs of particles and antiparticles, which effectively gives an acceleration.

We further provide an expression for the free energy, showing at infinite volume the thermodynamic Bethe ansatz (TBA) with Boltzmann-Maxwell statistics (see e.g.~\cite{doyon_lecture_2019}); and, remarkably, a similar structure {\em at finite volumes} generalising the recent result\footnote{We became aware of this after we obtained our results} in hard rods \cite{bulchandani2023modified} (we are not aware of any other examples). We then show that the GHD equation emerges in the large space-time limit, in the generality of arbitrary two-body shift. We confirm this by numerical simulations.

The models we introduce are different from most known classical integrable systems, whose dynamics are not of tracer type. They widely generalise hard-rod gases \cite{Spohn1991,ferrari2022hard}, and are closely related to the quantum Bethe ansatz and the gas of Bethe wave packets introduced recently \cite{doyon2023ab}. We refer to them as {\it semiclassical Bethe systems}. Some of the results presented here are proved rigorously in the separate paper~\cite{scb_long}.

{\bf The model.}--- 
Consider the $N$-particle classical  phase space with canonical coordinates $(\bm{y},\bm{\theta})\in\mathbb{R}^{2N}$, $\{y_i,\theta_j\}=\delta_{ij}$ -- which will be identified with asymptotic coordinates -- and the free-particle Hamiltonian $H(\bm{y},\bm{\theta}) = \sum_{i=1}^N \theta_i^2/2$.
Let $\psi(x,\theta)$ satisfy $\psi(-x,-\theta) = \psi(x,\theta)$ and $|x|\partial_x\psi(x,\theta) \to 0$ ($|x|\to\infty$). The generating function
\begin{align}\label{equ:generating_function}
    \Phi^{[\psi]}(\bm x,\bm\theta) =
	\sum_i x_i\theta_i
	+\frac{1}{2}\sum_{i, j} \psi(x_i-x_j,\theta_i-\theta_j),
\end{align}
induces a canonical transformation to the ``real" coordinates $(\bm{x},\bm{p})$ as $\bm y = \bm\nabla_{\theta} \Phi^{[\psi]}$, $\bm p = \bm\nabla_{x} \Phi^{[\psi]}$:
\begin{align}\label{cba1}
    y_i  &=
    x_i + \sum_{j\neq i}\partial_\theta\psi(x_i-x_j,\theta_i-\theta_j)\\ \label{cba2}
    p_i &= \theta_i + \sum_{j\neq i}
    \partial_x\psi(x_i-x_j,\theta_i-\theta_j)
\end{align}
where $\p_x$, resp.$\p_\theta$, means derivative with respect to the first, resp.~second, argument of $\psi(x,\theta)$. The Hamiltonian takes the form
\beq\label{Hpsi}
    H^{[\psi]}(\bm x,\bm p) = \sum_{i=1}^N \frc{\theta_i(\bm x,\bm p)^2}2=\sum_{i=1}^N \frc{p_i^2}2 + V^{[\psi]}(\bm x,\bm p)
\eeq
where $\theta_i(\bm x,\bm p)$ are obtained by solving \eqref{cba2} (see below) and the ``quasi-potential" $V^{[\psi]}(\bm x,\bm p)$ is defined by the second equation. The trajectories in phase space $t \mapsto (\bm{x}(t),\bm{p}(t))$ are induced from the free dynamics $\bm{y} \to \bm{y}(t) = \bm{y}+ \bm{\theta}t$ via the change of coordinates Eqs. \eqref{cba1} and  \eqref{cba2}. Note that there could be multiple trajectories that are admissible when $\p_x\p_\theta\psi(x,\theta)$ is negative.

Three statements hold: (i) The Hamiltonian \eqref{Hpsi} is Liouville integrable: there are $N$ independent conserved quantities, including the Hamiltonian, that Poisson commute with each other and are nice enough functions of phase space. Indeed a natural set is $Q_a=\sum_i \theta_i(\bm x,\bm p)^a = \sum_ip_i^a + V_a^{[\psi]}$, $a=1,\ldots N$.
(ii) The quasi-potentials are short-range. This is in the weak sense that whenever particles lie on well-separated intervals, $A_1,A_2\subset\R,\,{\rm dist}(A_1,A_2)\to\infty$, $I_1\cup I_2 = \{1,\ldots,N\}$, $\bm x_{I_i}\subset A_i$, then these do not interact, $V_a^{[\psi]}(\bm x,\bm p) \to \sum_i V_a^{[\psi]}(\bm x_{I_i},\bm p_{I_i})$. An important consequence is that one can define conserved densities $q_a(x)$ such that $Q_a = \int \dd x\,q_a(x)$, and with $\{q_a(x_1),q_{a'}(x_2)\}\to0$ for $x_i\in A_i$. Thus, in the limit of $N$ large for physically sensible finite-density distributions, conserved densities commute at large distances. (iii) The multi-particle scattering processes are {\em elastic} -- the sets of incoming and outoing momenta are the same -- and {\em factorise} into two-body scattering processes. The two-body scattering shift for incoming momenta $\theta_1,\theta_2$ is given by $\varphi(\theta_1-\theta_2)$ where $\varphi(\theta)
    = \lim_{x\to\infty}(\partial_\theta\psi(x,\theta)
	- \partial_\theta\psi(-x,\theta))$.
Factorised scattering means that in an $N$-body scattering event, outgoing particle $\theta$ is shifted, with respect to the straight trajectory of incoming particle $\theta$, by the sum $\omega(\theta) = \sum_{\theta'\neq \theta} \sgn(\theta'-\theta)\varphi(\theta-\theta')$ of the two-body shifts with all particles that it has crossed. Factorised scattering is a fundamental property of many-body integrable systems \cite{ZZ1979}. Here, because $\psi(x,\theta)$ becomes constant in $x$ as $x\to\pm\infty$, at long times $p_i\sim\theta_i$, and $x_i\sim y_i(t) +s_i^\pm$ with $s_i^+-s_i^-=\omega(\theta_i)$. Thus, $\theta_i$ are asymptotic momenta and $y_i$ are simply related to the impact parameters of the scattering process. Note that each particle $i$ has the same incoming and outgoing momentum $\theta_i$: thus it is a tracer for where the asymptotic momentum lies at finite times. We call this a {\em tracer dynamics}.

For any real symmetric $\varphi(\theta)$, we may choose $\psi(x,\theta) = f(x)\phi(\theta)$ where $\phi(\theta)=\int_0^\theta \dd\theta'\varphi(\theta')$ and $f(x)$ interpolates between $-1/2$ and $1/2$, e.g.~$f(x) = \tfrac{x}{2\sqrt{x^2+\alpha^2}}$ for some $\alpha\neq 0$. Therefore, this is a new family of classical Liouville integrable, short-range, factorised-scattering models, covering {\em a large class of two-body shift functions}. Only a few shift functions are known to date to correspond to classical integrable models, hence this is a large extension. Note that fixing the scattering does not fix the dynamics: there is still freedom in $\psi(x,\theta)$. In \cite{scb_long}, assuming that $\psi$ is twice continuously differentiable, and that $\p_\theta\p_x\psi(x,\theta)\geq 0$ (thus $\varphi(\theta)\geq 0$) along with a finite-range condition, we rigorously show the statements above, and we show that \eqref{cba1}, \eqref{cba2} have unique continuously differentiable solutions. We believe much of this stays true under weaker conditions; but uniqueness can be broken, leading to important physical effects that we discuss below.

By a judicious choice of $\psi(x,\theta)$ one can reproduce the hard rod dynamics \cite{Spohn1991,cardy2022t,ferrari2022hard}, see the Supplemental Material \cite{SM}. The generating function \eqref{equ:generating_function} has the structure of a phase $\Phi$ for Bethe wave functions  $\Psi = e^{\ri \Phi}$, with Bethe roots $\theta_i$ and dynamics from semiclassical arguments: $\bm p = \bm\nabla_{x} \Phi$ are physical momenta, and $\bm y = \bm\nabla_{\theta} \Phi$ evolve trivially. In the Lieb-Liniger model, gases of wave packets are indeed described by $\psi_{\rm ll}(x,\theta)=\frc12\sgn(x)\phi(\theta)$ with $\phi(\theta)$ the quantum scattering phase \cite{doyon2023ab}, and we expect a similar relation for most quantum many-body integrable systems. 

\begin{figure}[!h]
    \centering
    \includegraphics[width=\linewidth]{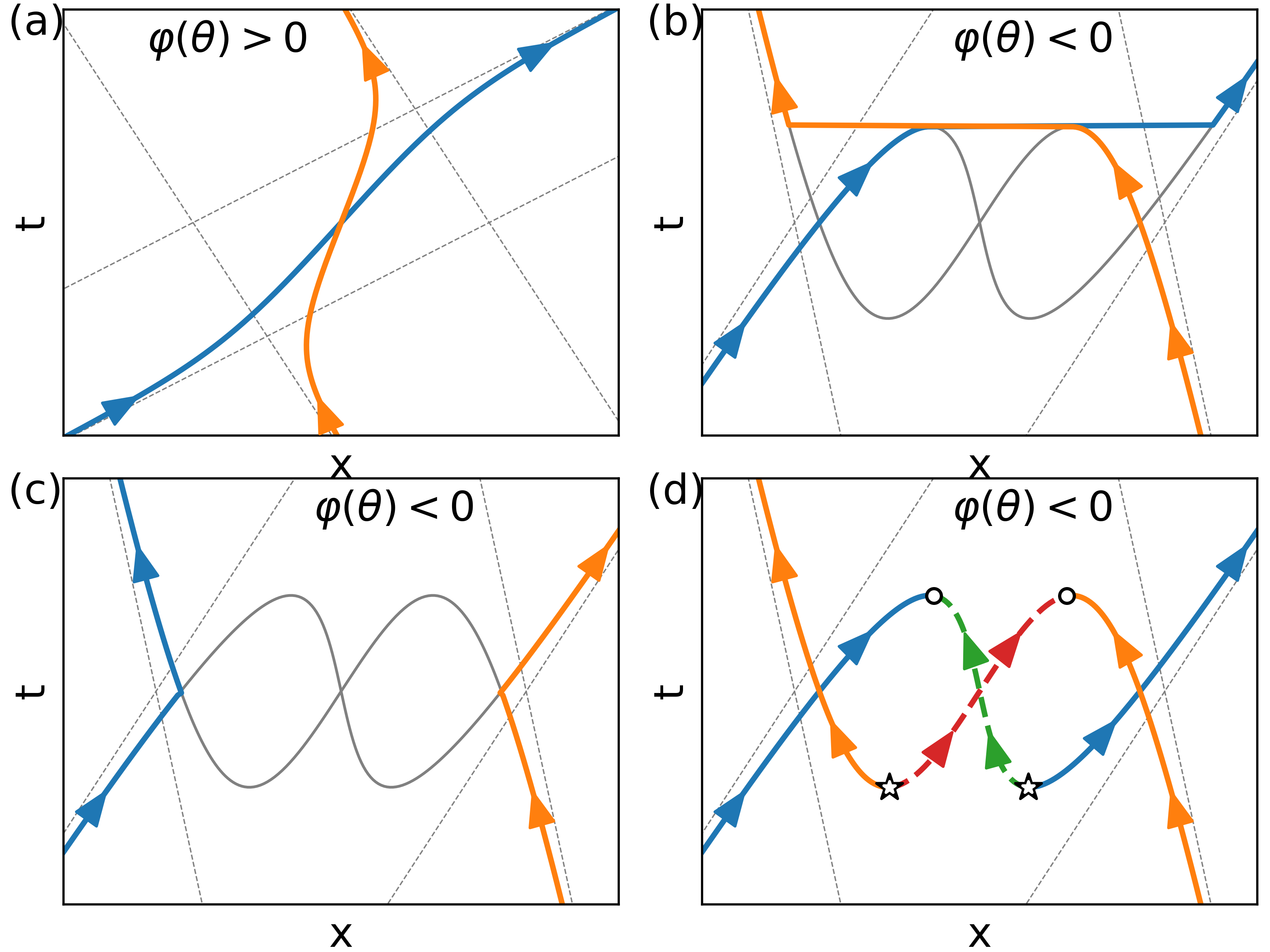}
    \caption{Trajectories of two particles during scattering for (a) $\varphi(\theta)>0$ and (b), (c), (d) $\varphi(\theta)<0$. The thin dashed lines are the asymptotic trajectories. We show three interpretations of the multivalued solutions for $\varphi(\theta)<0$: (b) Particles jump instantaneously. (c) Particles are relabelled during collisions. (d) Spontaneous creation of a particle-antiparticle pairs: starting from the bottom, at a certain time (marked by a star) the blue and orange particles are close enough to allow for two pair creations (there are 3 solutions: two blue-orange (outer), one red-green (inner)). Later one particle of each pair (the red and green) each annihilates one of the original particles (marked by a circle), leaving the new blue and orange particles as outgoing. We used $\psi(x,\theta) = \tfrac{x}{\sqrt{x^2+\alpha^2}}2\arctan{\tfrac{\theta}{c}}$, with $c=0.5,\alpha = 0.4$ for (a) and $c=-1,\alpha = 1$ for (b), (c), (d).}
    \label{fig:two_particles_interpretation}
\end{figure}
{\bf Microscopic dynamics.}--- The effect of the interaction can be seen as a particle-dependent, dynamical change of metric from $x$- to $y$-space where it is free: the change of infinitesimal length is $\dd y_i = K_i(x_i)\dd x_i$ where $K_i(x) = \big(1 + \sum_{j\neq i}\p_\theta\p_x\psi(x-x_j,\theta_i-\theta_j)\big)$ measures the effective ``free" space. It can best be pictured in the two-particle case, see Fig.~\ref{fig:two_particles_interpretation} and \cite{SM}. For $\varphi(\theta) > 0$ (e.g.~$\psi_{\rm ll}$) particles slow down during scattering, giving an effective backwards displacement  (Fig.~\ref{fig:two_particles_interpretation}a) interpreted as the presence of additional, hidden space where particles must travel. If $\psi(x,\theta)=\frc12\sgn(x)\phi(\theta)$ for some $\phi(\theta)$, they ``stick" and acquire an internal clock that accounts for this extra space at collisions \cite{doyon2023ab}.  For $\varphi(\theta) < 0$, \eqref{cba1} does not necessarily have a unique solution. In case it does, particles speed up, giving an effective forward displacement and a reduction of effective space; hard rods of positive lengths are a limiting case, where the displacement -- which traces the momentum being transferred -- is instantaneous. But for most choices of $\psi(x,\theta)$ with $\varphi(\theta)<0$, solutions to \eqref{cba1} can become multivalued. Then trajectories appear to go backward in time, and the generating function \eqref{equ:generating_function}, although locally inducing a canonical transformation, globally does not on the standard phase space.
One may consider three ``regularisations", without affecting the large-scale physics, all implementing a reduction of effective space. (i) Choose any branch; e.g.~follow one branch until it disappears, then jump to another branch (Fig.~\ref{fig:two_particles_interpretation}b). This is similar to the flea gas \cite{PhysRevLett.120.045301} (we do not know if there exist choices of $\psi(x,\theta)$ and branch that would exactly reproduce the flea gas algorithm). But, like for the flea gas, this regularisation is not Hamiltonian nor time reversible. (ii) Using the ``hard-core" picture \cite{cardy2022t}, relabel particles at the first collision (Fig.~\ref{fig:two_particles_interpretation}c). This gives a time-reversible dynamics (no longer a tracer dynamics), as long as such collisions always appear before any time-backward parts of trajectories; this re-labelling gives the rods in the hard-rod case. (iii) Inspired by Feynman's picture, interpret time-backward parts of trajectories as antiparticles (Fig.~\ref{fig:two_particles_interpretation}d). The proximity of a particle (say orange in the figure) occasions a spontaneous particle-antiparticle pair creation (blue and green); the antiparticle (green) later annihilates with the incoming particle (blue) leaving the created particle  (also blue) as outgoing physical particle. This is time-symmetric, and we believe it might define a canonical flow on the `Fock phase-space' $\mathcal{F} = \bigoplus_g \mathbb{R}^{2Ng}$ which admits an arbitrary number $g$ of solutions to \eqref{cba1}; however this would need to be investigated further. For smooth $\psi(x,\theta)$, multivaluedness can always be interpreted in this way, as the equations \eqref{cba1}, \eqref{cba2}, for $\boldsymbol x,\boldsymbol p, t$, define smooth curves in $\R^{2N+1}$. We now argue that this picture naturally arises from \ttbar-deformations.

{\bf \ttbar-deformations.}---
Generalized \ttbar-deformations, as proposed in \cite{10.21468/SciPostPhys.13.3.072}, are obtained as flows of Hamiltonians $H^{(\lambda)}\to H^{(\lambda)}+\delta H^{(\lambda)}$ parametrized by $\lambda$:
\begin{align}
    \label{genTTbarsummary}
	\delta H^{(\lambda)}&= \delta\lambda\int \dd\theta\dd\alpha
	\dd x\dd x' \, w(x-x',\theta-\alpha) \n
 &\quad\times\big( q^{(\lambda)}_{\theta}(x)j^{(\lambda)}_{\alpha}(x')
	-j^{(\lambda)}_{\theta}(x)q^{(\lambda)}_{\alpha}(x')\big),
\end{align}
with $w(x,\theta)$ some deformation function. Here $q^{(\lambda)}_{\theta}(x)$ and $ j^{(\lambda)}_{\theta}(x)$ are the charge densities and currents, with continuity equation $\p_tq^{(\lambda)}_{\theta}+\p_xj^{(\lambda)}_{\theta}=0$, associated to the charge $Q^{(\lambda)}_\theta=\sum_i\delta(\theta-\theta^{(\lambda)}_i)$ that measures the density of asymptotic momenta at $\theta$ in the deformed system. It turns out that if $H$ is an integrable tracer dynamics, then so is $H^{(\lambda)}$, and $q^{(\lambda)}_{\theta}(x)$ and $ j^{(\lambda)}_{\theta}(x)$ exist and are short-range.
Remarkably, starting from a system of free particles $H^{(0)} = \sum_i p_i^2/2$, the deformed Hamiltonian $H^{(\lambda)}$ is nothing else but $H^{[\psi^\lambda]}$, Eq.~\eqref{Hpsi}, with $\psi^\lambda(x,\theta) = 
	\lambda \int_{-\infty}^\infty \dd x'\,\sgn(x-x')w(x',\theta)$.
This generalises the mass-momentum deformation yielding hard rods \cite{cardy2022t}, see \cite{SM}. The semiclassical Bethe systems are the first concrete example of generalized \ttbar-deformations. We have rigorous proofs of these statements \cite{scb_long} under conditions guaranteeing invertibility of \eqref{cba1}, \eqref{cba2}, where $H^{[\psi^\lambda]}$ can be constructed on the standard phase space.

Going further, here we make the crucial observation that the relation  Eq.~\eqref{cba1}, be it invertible or not, is still the correct \ttbar-deformation of the impact-parameter-position relation $y_i=x_i$. Indeed, \ttbar-deformations \eqref{genTTbarsummary} can be obtained as {\em canonical flows} \cite{Pozsgay2020,10.21468/SciPostPhys.9.5.078,cardy2022t,scb_long}, and Eq.~\eqref{cba1} arises directly from applying this flow. Thus, {\em multivaluedness may appear along the \ttbar-flow}, and pair creation/annihilation processes occur (Fig.~\ref{fig:two_particles_interpretation}d), as claimed; see \cite{SM}. In all cases, the dynamics remains local.

{\bf Thermodynamics.}--- 
Let us consider the generalised Gibbs ensembles \cite{PhysRevB.95.115128}, with Boltzmann weights $e^{-\sum_{a}\beta_a Q_a}$. We take more generally $(x,\theta)$-dependent Lagrange parameters varying on scale $L$, with $e^{-\int \dd x\dd\theta\,\beta(x/L,\theta) q_\theta(x)}=e^{-\sum_i\beta(x_i/L,\theta_i)}$. In \cite{scb_long} we show, using methods of graph theory \cite{10.1007/978-981-13-2179-5_6} and under certain further assumptions, that the free energy density
\begin{align}
    \label{fLsummary}
	f_L &= -\frac{1}{L} \log \sum_{N=0}^\infty
	\int \frac{\dd^Nx \dd^Np}{N!}e^{-\sum_i \beta(x_i/L,\theta_i(\bm x,\bm p))}
\end{align}
is finite and given by $f_L = -(2\pi L)^{-1}\int\dd x\dd\theta\,e^{-\varepsilon_L(x,\theta)}$ where the pseudo-energy $\varepsilon_L(x,\theta)$ satisfies
\begin{equation}\label{eq:finite_pseudo}
    \varepsilon_L(x,\theta)=\beta(\tfrac{x}L,\theta)-\int\frac{\dd x'\dd \theta'}{2\pi}\partial_\theta\partial_x\psi(x-x',\theta-\theta')e^{-\varepsilon_L(x',\theta')}.
\end{equation}
This is a TBA-like equation for Maxwell-Boltzmann statistics (see e.g.~\cite{doyon_lecture_2019}). The TBA is well known from quantum \cite{yangyang,takahashi_1999,Zamolodchikov1990} and classical \cite{bastianello2018generalized,spohn_generalized_2019,doyon_generalised_2019,koch2022generalized,10.21468/SciPostPhys.15.4.140} integrability, at infinite volumes. It is striking that even at {\em finite volumes} the free energy possesses a TBA structure; the only result we are aware about this is for (positive-length) hard rods \cite{percus1976,bulchandani2023modified}. We postulate that the finite-volume free energy in interacting quantum and classical integrable systems is given by Eq.~\eqref{eq:finite_pseudo} under an appropriate choice of $\partial_\theta\partial_x\psi(x,\theta)$ reproducing the scattering shift $\varphi(\theta)$. The infinite-volume limit of Eq.~\eqref{eq:finite_pseudo} yields the expected TBA equation $\varepsilon(\b x,\theta)=\lim_{L\to\infty}\varepsilon_L(L\b x,\theta)=\beta(\b x,\theta)-\int\frac{\dd \theta'}{2\pi}\varphi(\theta-\theta')e^{-\varepsilon(\b x,\theta')}$, from which $\lim_{L\to\infty} f_L$ follows.  Thus the thermodynamics of the semiclassical Bethe systems is described by the standard machinery of TBA, including its ``local density approximation".

The free energy gives thermodynamic averages and fluctuations of conserved quantities. Interestingly, we also have the exact thermodynamic average $\rho_{\rm phys}(p) = n(\theta(p))/(2\pi)$ for the {\em physical momentum distribution} $L^{-1}\sum_i\delta(p-p_i)$. Here $n(\theta)=e^{-\varep(\theta)}$ is the occupation function and $\theta(p)$ is the inverse of the ``Dressed momentum" function $p^{\rm Dr}(\theta)$. The latter is known to be the physical momentum of an excitation at Bethe root $\theta$ in Bethe ansatz systems, and is fixed by TBA equations. See \cite{SM}. As physical momenta of particles change throughout their trajectories, $\rho_{\rm phys}(p)$ is a quantity that is typically hard to access in integrable models; this is the first exact expression that we are aware of.

{\bf GHD.}--- We provide a heuristic argument for GHD to emerge in the hydrodynamic limit, paralleling Ref.~\cite{doyon2023ab}; other techniques \cite{Boldrighini1983} should give rigorous results.

We take macroscopic space and time, $x= L\b x,\,t = L\b t$ ($\b x$, $\b t$ finite, $L\to\infty$), with scaled coordinates $\b x_i(\b t) = x_i(t)/L$ and $\b y_i = y_i/L$. The empirical density $\rho_{\rm e}(\theta,\b x,\b t) = L^{-1}\sum_i \delta(\b x-\b x_i(\b t))\delta (\theta-\theta_i)$,
is assumed to converge ``weakly": $\rho_{\rm e}(\theta,\b x,\b t) \to \rho_{\rm p}(\theta,\b x,\b t)$. Clearly,
\beq\label{contrhoe}
	\p_{\b t} \rho_{\rm e}(\theta,\b x,\b t) +
	\p_{\b x} \Big(L^{-1} \sum_i \dot{\b x}_i
	\delta(\b x-\b x_i)\delta(\theta-\theta_i)\Big) = 0.
\eeq
Re-writing Eq.~\eqref{cba1} as $\b x_i(\b t) =\b y_i + \theta_i \b t
	- \frc1L\sum_{j\neq i} \partial_\theta\psi(L(\b x_i(\b t) - \b x_j(\b t)),\theta_i-\theta_j)$,
we assume that, for every $i$, there is a fraction of particles $j$ that tend to 1 as $L\to\infty$ such that $\b x_i(\b t) - \b x_j(\b t)>0$. Then, we can replace $\partial_\theta\psi(L\b x,\theta) \to \psi^{\sgn(\b x)}\varphi(\theta)$ where $\psi^+ - \psi^- = 1$. Taking the $\b t$-derivative, we find
\beq\label{dotbarxi}
	\dot{\b x}_i = \theta_i - \frc1L\sum_{j\neq i}
	\delta(\b x_i-\b x_j)\varphi(\theta_i-\theta_j)
	(\dot{\b x}_i - \dot{\b x}_j) \qquad (L\to\infty).
\eeq
Making the ansatz $\dot {\b x}_i = f(\b x_i,\theta_i)$, the second term on the right-hand side is $-
	\int \dd\theta\,\rho_{\rm e}(\theta,\b x_i,\b t)
	\varphi(\theta_i-\theta)(f(\b x_i,\theta_i) - f(\b x_i,\theta))$.
Thus $f(\b x,\theta) = v^{\rm eff}_{[\rho_{\rm e}(\cdot,\b x,\b t)]}(\theta)$ solves
\beq
	v^{\rm eff}_{[\rho]}(\theta) = \theta -
	\int \dd\theta'\,\rho(\theta')
	\varphi(\theta-\theta')
	(v^{\rm eff}_{[\rho]}(\theta) - v^{\rm eff}_{[\rho]}(\theta')).
\eeq
This is exactly the equation for the effective velocity in GHD \cite{PhysRevX.6.041065,PhysRevLett.117.207201}. Putting this into Eq.~\eqref{contrhoe} and taking the limit $L\to\infty$ we obtain the GHD equation,
\beq
	\p_{\b t} \rho_{\rm p}(\theta,\b x,\b t) +
	\p_{\b x} \big(v^{\rm eff}_{[\rho_{\rm p}]}(\theta,\b x,\b t) \rho_{\rm p}(\theta,\b x,\b t)\big)=0.
\eeq
Thus, we have shown that, at large scales, the semiclassical Bethe system satisfies the GHD equation. This is not rigorous -- in particular, in Eq.~\eqref{dotbarxi} one would need to use a regularization of the delta-function. We  give an alternative derivation of GHD in \cite{SM}, based on the fact that the metric change $\dd y_i = K_i(x_i)\dd x_i$ converges to the GHD change of metric $K_i(x)\to 2\pi \rho_{\rm s}(\theta_i,x)$ determined by the ``space" or ``total" density $\rho_{\rm s}(\theta,x)$ \cite{Doyon2018}.

\begin{figure}[!h]
    \centering
    \includegraphics[width=\linewidth]{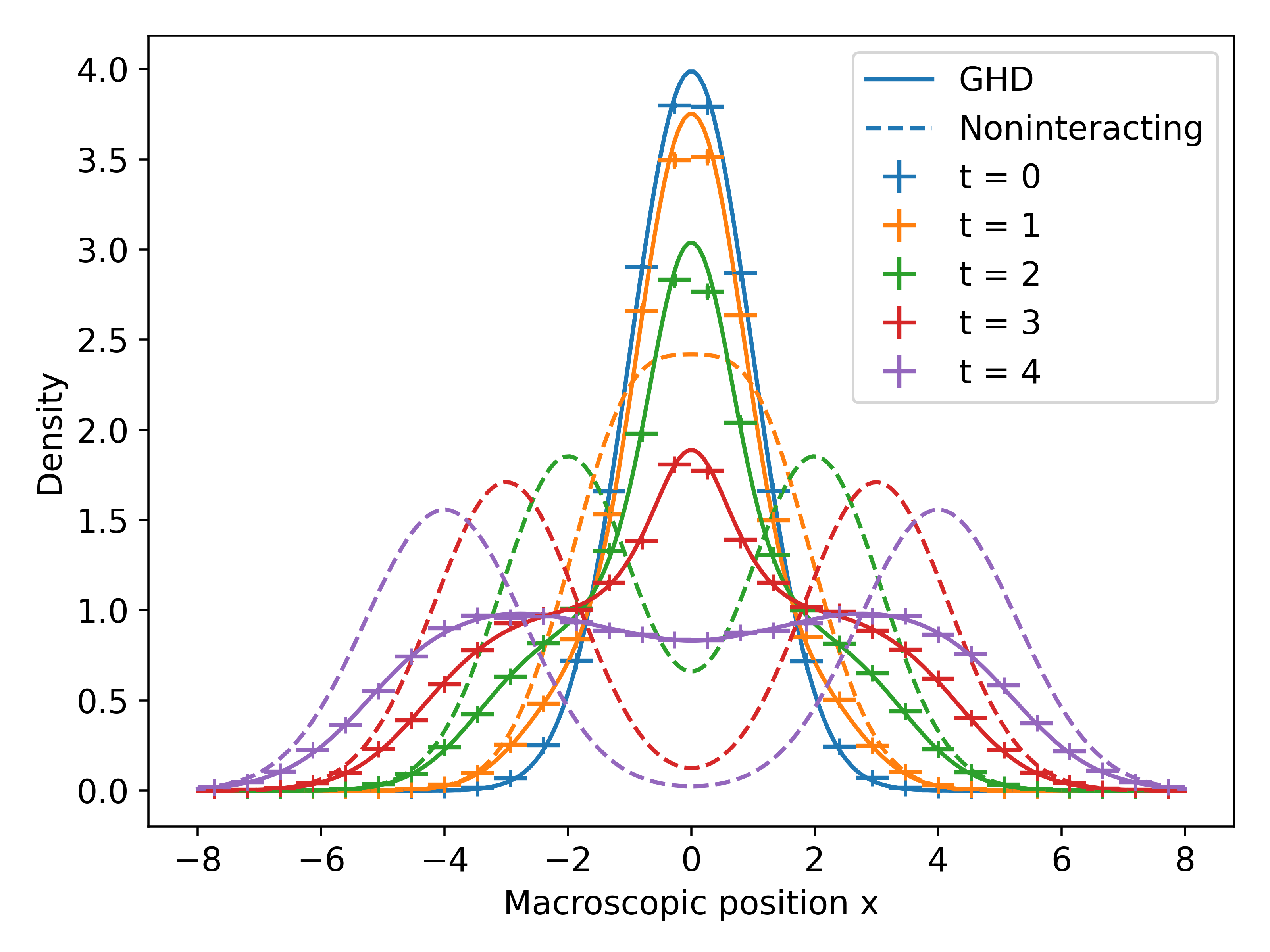}
    \caption{Evolution of the microscopic particle density (data points) and that from the GHD solution (solid lines) for different times. The initial state is  $\rho_{\rm p}(\theta,x,0) = \frac{25}{2\pi}e^{-x^2/2} \left(e^{-25(\theta-1)^2/2}+e^{-25(\theta+1)^2/2}\right)$. We use $\psi(x,\theta) = \tfrac{x}{2\sqrt{x^2+\alpha^2}}\phi(\theta)$, where $\alpha = 0.1$ and $\phi(\theta) = 2\arctan{\tfrac{\theta}{c}}$. The microscopic results were averaged over $100$ samples, each starting from a randomly generated initial state ($L=300$, each sample contains $N \approx 3000$ particles). The error bars indicate the standard deviation and the bin size used for computing the density of particles. We also give the density for non-interacting particles (dashed lines) for comparision.}
    \label{fig:GHD}
\end{figure}

We numerically demonstrate that the GHD equation correctly captures the large-scale behaviour in an explicit example, see Fig.~\ref{fig:GHD}. For illustration, we use the phase shift from the quantum Lieb-Liniger model, but with an initial state that breaks the maximal fermionic occupation allowed by quantum mechanics (the maximal density of particle per state is $6.264 > 1$). This initial state is nevertheless realizable, and its hydrodynamics makes sense, as indeed it is realised by a semiclassical Bethe system. The details of the numerical simulations can be found in \cite{SM}. Compared to the evolution of non-interacting particles the expansion of the interacting particles is much slower, which is in line with the intuitive meaning of a positive phase-shift as an effective time-delay during the scattering of two particles.

{\bf Conclusions.}--- We introduced a new class of classical integrable models, dubbed semiclassical Bethe systems for their relation with the quantum Bethe ansatz, obtained as generalized \ttbar-deformations of classical noninteracting particles. In these systems, each particle is a ``tracer": it has the same incoming and outgoing momentum. The class is parametrised by a function determining the microscopic dynamics, and displays factorised scattering with a {\em (largely) arbitrary two-body shift}, including those found in many quantum integrable models. The microscopic dynamics displays special features including pair creations / annihilations; the thermodynamics in finite volumes surprisingly takes a form akin to the thermodynamic Bethe ansatz (TBA), reducing to the TBA at infinite volume; the distribution in physical phase space can be evaluated exactly; and the large-scale dynamics is described by GHD, and therefore {\em identical} to that of any quantum/classical integrable system with the same chosen two-body shift. We conjecture that, with short-range interaction, the agreement persists at higher orders: the models should encode the universal hydrodynamic expansion of classical many-body integrability, as corrections due to specific interactions should be exponentially subleading. For instance, particles' positions in our models should approximate well the spatial distribution of solitons in dense soliton gases, something that can be useful for initial state preparation.

It would be interesting to construct the full particle-non-conserving Hamiltonian description of trajectories \eqref{cba1}, \eqref{cba2} with negative shifts $\varphi(\theta)<0$, Fig.~\ref{fig:two_particles_interpretation}d. Finding the full integrability structure of our models, perhaps connecting with sine-Gordon soliton trajectories \cite{babelon1993sine,babelon1996quantization,babelon1997form}, would be interesting, as would quantising our models, perhaps in the spirit of \cite{doyon2023ab} (integrability of generalized \ttbar-deformed systems is established \cite{10.21468/SciPostPhys.13.3.072});
the notion of pair creation / annihilation may play an important role. Finally, adding an external potential is possible, and we anticipate that rigorous proofs of the emergence of the GHD equation can be obtained following ideas in the hard-rod case \cite{Boldrighini1983,ferrari2022hard}. This might also shed light on GHD beyond the Euler scale.

{\bf Acknowledgements.}---
We are grateful to Olalla Castro Alvaredo for discussions, and to Joseph Durnin for previous collaboration on related subjects. FH acknowledges funding from the faculty of Natural, Mathematical \& Engineering Sciences at King's College London. BD was supported by the Engineering and Physical Sciences Research Council (EPSRC) under grant EP/W010194/1. The numerical simulations were done using the CREATE cluster~\cite{CREATE}.

\bibliographystyle{apsrev4-1}
\bibliography{bib.bib}

\end{document}



\title{Supplemental material for\\ ``New classical integrable systems from generalized \ttbar-deformations"}

 \author{Benjamin Doyon}
\affiliation
{Department of Mathematics, King’s College London, Strand, London WC2R 2LS, U.K.}
\author{Friedrich H\"ubner}
\affiliation{Department of Mathematics, King’s College London, Strand, London WC2R 2LS, U.K.}
\author{Takato Yoshimura}
\affiliation{All Souls College, Oxford OX1 4AL, U.K.}
\affiliation{Rudolf Peierls Centre for Theoretical Physics, University of Oxford,
1 Keble Road, Oxford OX1 3NP, U.K.}

\maketitle

In this supplemental material, we refer to equations in the main text as \eqrefmt{{\em x}} where {\em x} is the equation number.

\section{Two-body scattering in the semiclassical Bethe system}
For two particles the evolution decouples in the trivial center of mass motion of $x_1(t)+x_2(t)$ and the motion of the relative coordinate $x_{12}(t) = x_1(t)-x_2(t)$:
\begin{align}
    y_{12}+\theta_{12}t = x_{12}(t) + 2 \partial_\theta\psi(x_{12}(t),\theta_{12}) := f(x_{12}(t))
\end{align}
Note that $f(x)$ is a one-dimensional function which satisfies $f(x\to\pm \infty) \to \pm \infty$ since $\psi_\theta(x,\theta)$ is bounded. Therefore for any time $t$ a solution $x_{12}(t)$ will always exist. If $f(x)$ is monotone increasing, i.e. $f'(x)=1+2 \psi_{x\theta}(x_{12}(t),\theta_{12}) > 0$ then this solution will also be unique for all times, otherwise not. This gives a bound $\partial_{x}\partial_\theta\psi(x,\theta)>-\tfrac{1}{2}$, which is automatically satisfied for positive $\partial_{x}\partial_\theta\psi(x,\theta)(x,\theta)$. Thus, given a positive phase shift $\varphi(\theta)$ it is easy to find a $\psi(x,\theta)$ that gives rise to unique trajectories, for instance $\psi(x,\theta) = \frac{x}{2\sqrt{x^2+\alpha^2}}\phi(\theta)$, where $\alpha>0$ is an arbitrary real number and $\phi'(\theta) = \varphi(\theta)$. On the other hand, if $\varphi(\theta)$ is negative, then non-uniqueness can easily happen if $\partial_{x}\partial_\theta\psi(x,\theta)$ is too large. For instance, consider again the implementation $\psi(x,\theta) = \frac{x}{2\sqrt{x^2+\alpha^2}}\phi(\theta)$: The trajectories are only unique if $\alpha > |\varphi(\theta_{12})|$, i.e. if the interaction region is broad. In fact, for fixed particle number $N$ and bounded $\varphi(\theta)$, it is always possible to choose a $\psi(x,\theta)$ with a sufficiently large interaction region, s.t. the trajectories are unique.

\begin{figure}
    \centering
    \includegraphics[width=\linewidth]{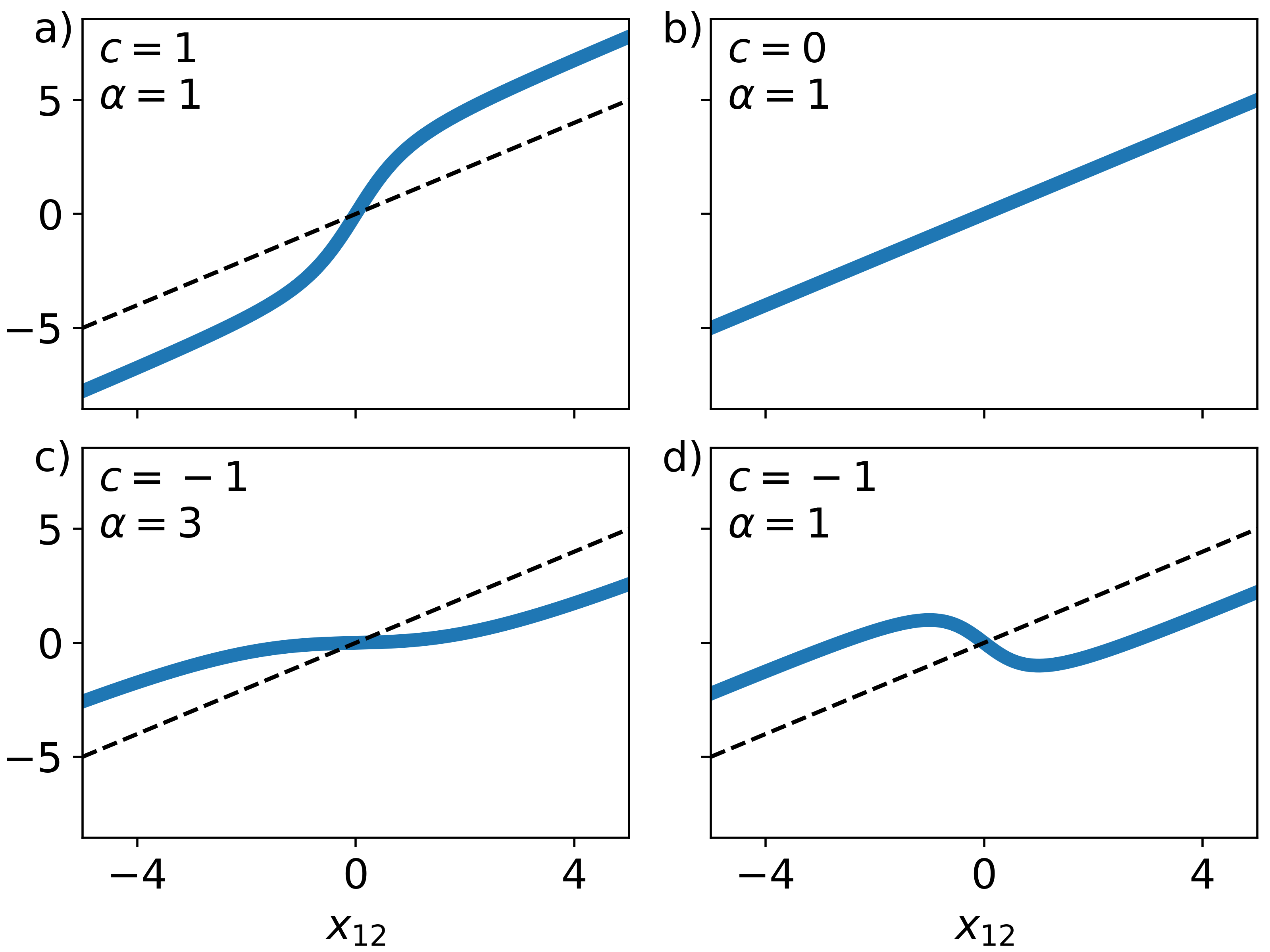}
    \caption{Relation of the relative particle positions in free space $y_{12} = y_1-y_2$ and deformed space $x_{12} = x_1-x_2$ for different $c$ and $\alpha$ (in this case $\partial_{\theta}\psi(x,\theta) = \frac{x}{\sqrt{x^2+\alpha^2}}\varphi(\theta)$, where $\varphi(\theta) = \frac{2c}{\sqrt{c^2+\theta^2}}$ is the Lieb-Liniger phase shift). The free particle solution b) for $c=0$ is given in the other plots for comparison (dashed line). For a) positive $c>0$ or c) large $\alpha$ the map is invertible. If $c<0$ and $\alpha$ too small then d) the map is non-invertible. This leads to the non-well definedness of particles during scattering.}
    \label{fig:two_particles}
\end{figure}

If $\alpha$ is too small the inverse function of $f$ becomes multivalued. In this case the trajectory becomes non-unique during scattering (see Fig. \ref{fig:two_particles}).

\section{Specialisation to hard rods}
\begin{figure}[!h]
    \centering
    \includegraphics[width=\linewidth]{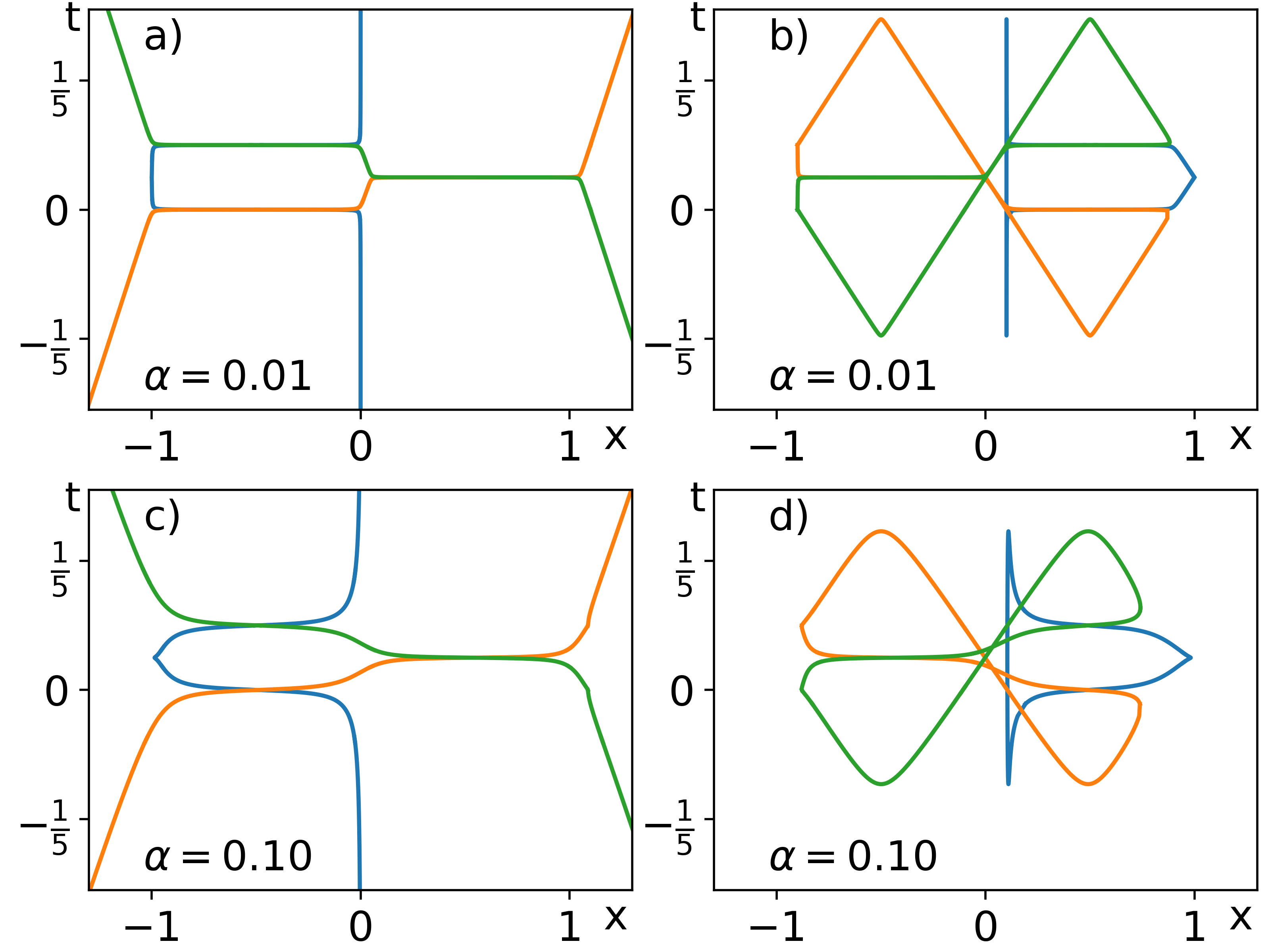}
    \caption{Trajectories of three $\lambda=1$ hard rod like particles during scattering obtained by numerically solving~\eqrefmt{2}. In order to have smooth behaviours, we regularize $\psi_{\rm hr,+}(x,\theta)$ as follows $\psi(x,\theta) = \tfrac{1}{4}(\sqrt{(x^2-d^2)^2+4\alpha^2}-\sqrt{(x^2+d^2)^2+4\alpha^2})$, which reduces to $\psi_{\rm hr,+}(x,\theta)$ as $\alpha \to 0$: a), b) depict the trajectories close to the hard rod limit while c), d) give smoother versions for comparison. The three particles are fixed at $t=0$ by $y_1=0,\theta_1=0$ (blue), $y_2=0,\theta_2=1$ (orange) and $y_3=0.1,\theta_1=-1$ (green). For both $\alpha$ we find two solutions, the ``normal sector" a), c) and the ``vacuum loops" b), d).}
    \label{fig:three_particles}
\end{figure}
For hard rods of positive length $\lambda$ \cite{Spohn1991}, one may take the continuous piecewise linear $\psi_{\rm hr,+}(x,\theta) = \{\frc\lambda2\;(x<-\lambda),\ -x/2\;(|x|<\lambda),\  -\frc\lambda2\;(x>\lambda)\}\times \theta$. The particles then trace the hard rods' momenta. Indeed, this guarantees that whenever particles' positions are separated by distances greater then $\lambda$, they move freely, while the first pair of particles coming to a distance $\lambda$ experiences an instantaneous motion pass each other, effectively exchanging their positions. As almost surely (with respect to physically sensible distributions of positions and momenta) only at most one pair of particles at a time come to a distance $\lambda$, only at most one two-body process occurs at a time. This indeed reproduces the hard-rods dynamics. This should be seen as the limit $\ep\to0$ of the choice $\psi_{\rm hr,+}^\ep(x,\theta) = \{\frc\lambda2\;(x<-(1+\ep)\lambda),\ -\frc{x}{2(1+\ep)}\;(|x|<\frc\lambda2),\  -\frc\lambda2\;(x>(1+\ep)\lambda\}\times \theta$. This choice gives invertible Eqs.~\eqrefmt{2}, \eqrefmt{3} on all configurations where at most one pair of particles are a distance smaller than $(1+\ep)\lambda$, and as $\ep\to0$, at a collision of a pair of particles the speeds of the particles tend to infinity, giving, in the limit, the instantaneous exchange of their positions.

For hard rods of negative length  $-\lambda$ \cite{cardy2022t,ferrari2022hard} a natural choice would be $\psi_{\rm hr,-}(x,\theta) = \frc12\sgn(x)\lambda\theta$. In fact, the standard negative-length hard-rod model  implements the negative lengths by allowing for rods' centers to cross each other -- thus being in the ``wrong order" -- up to a distance $\lambda$ before changing direction; while here particles simply stick with each other for a time in order to implement the same overall scattering shift \cite{doyon2023ab}. As the motion of the particles stuck with each other simply takes the average momentum \cite{doyon2023ab}, in fact, this model represents the {\em center-of-mass position} of any group of rods in the wrong order; the microscopic dynamics is therefore slightly different.

In the case of positive-length rods, in fact there is one subtlety concerning the dynamics with $\psi_{\rm hr,+}$. Let us refer to the ``normal sector" as the set of configurations under the constraint that all but at most one pair of particles are a distance strictly greater than $\lambda$ from each other (or $(1+\ep)\lambda$ in the regularised version). Although the solutions to Eqs.~\eqrefmt{2}, \eqrefmt{3} in  are indeed unique in the normal sector, {\em there are in general other solutions} where many particles are near to each other. These ``ghost solutions" are not required for a well-defined dynamics when starting with configurations where all particles are far enough from each other, because, as explained above, almost surely the normal sector only is explored over time. However, if admitting all solutions, as per our general particle-production process, these solutions should be considered; they produce ``vacuum" loops. Such loops in fact allow us to {\em define the hard-rod dynamics, for positive rod lengths $\lambda$, with rods being closer than a distance $\lambda$}, thus in particular at densities higher than $1/\lambda$.

Therefore, we see that the hard-rod model with positive rod lengths is that given by the restriction to the normal sector of the semiclassical Bethe systems with $\psi_{\rm hr,+}$; this sector is almost surely  invariant under the dynamics, hence a consistent restriction.

We note that the choice $w(x,\theta)=\frc12\delta(x)\theta$ in \eqrefmt{5}, for the generalised \ttbar-deformation, corresponds to the mass-momentum deformation, argued in \cite{cardy2022t} to give rise to the hard rods. From our analysis, we see that a number of subtleties arise.

First, in \cite{cardy2022t}, it is the analytical continuation in $\lambda$, from positive rod lengths, that was argued to give negative-length rods. Here, we see that, as per our discussion above, the direct deformation towards negative rod lengths gives $\psi_{\rm hr,-}$, the ``center-of-mass" version where for groups of rods in the wrong order, we only trace the momentum of the center-of-mass position. Second, the choice $w(x,\theta)=\frc12\delta(x)\theta$, in the direction of the \ttbar-flow giving positive rod lengths, in fact only gives the true positive-length hard rods under the hard-core regularisation where exchanges are made at first collisions (see the main text); this was indeed the picture taken in \cite{cardy2022t}. Here, we see that one must choose a $\lambda$-dependent $w_{\rm hr,+}(x,\theta) = \frc1{4\lambda}\Theta(|x|-\lambda)$ in order for \ttbar-deformation to give the semiclassical Bethe system with $\psi_{\rm hr,+}$ that directly represent, when restricting to the normal sector, the tracers of hard rods' momenta (the go-through picture).

Finally, the specialisation of the finite-volume TBA equation \eqrefmt{7} to the center-of-mass negative-length hard rods, with $\psi_{\rm hr,-}$, is
\begin{equation}
    \varepsilon_L(x,\theta)=\beta(\tfrac{x}L,\theta)-\lambda\int\frac{\dd \theta'}{2\pi}e^{-\varepsilon_L(x,\theta')}.
\end{equation}
It is interesting that we find no correction from the local structure at all: the same TBA equation arises at infinite volume. As \eqrefmt{7} is derived in \cite{doyon2023generalised} under assumptions that are not satisfied for the positive-length hard rods, $\psi_{\rm hr,+}$, its specialisation to this case is not expected to give the correct answer (this turns out to be of a similar form to that of \cite{percus1976,bulchandani2023modified}, but sligthly different).



\section{Details on the numerical simulations}
In the letter we simulate the Euler scale time evolution of a macrosopic initial state described by:
\begin{align}
    \rho_0(x,\theta) = \frac{25}{2\pi}e^{-x^2/2} \left(e^{-25(\theta-1)^2/2}+e^{-25(\theta+1)^2/2}\right),\label{equ:simulation_rho0}
\end{align}
using both the underlying microscopic particle model and the GHD equation. In the following we provide details on both simulations.

\subsection{Particle simulations}
\subsubsection{Generating the initial state}
In order to generate an initial state whose quasi-particle density corresponds to Eq. \eqref{equ:simulation_rho0}, we use an adaption of an algorithm that is used to generate hard rods initial states~\cite{10.21468/SciPostPhys.15.4.136}. The algorithm first generates the positions of particles using their total density $\bar{\rho}_0(x) = \int\dd{\theta} \rho_0(x,\theta) = \frac{10}{\sqrt{2\pi}}e^{-x^2/2}$ and then subsequently randomly chooses their initial $\theta$ according to the double Gaussian momentum distribution $f(\theta) = \frac{\rho_0(x,p)}{\bar{\rho}_0(x)} = \frac{1}{5\sqrt{2\pi}}\left(e^{-25(\theta-1)^2/2}+e^{-25(\theta+1)^2/2}\right)$.

The procedure for generating the particle positions is as follows: Fix the location of the first particle (we choose $x_1 = 0$). The locations of all the particles for $x>0$ are generated consecutively using the following rule:
\begin{align}
    x_{n+1} = x_n + \eta_n,
\end{align}
where $\eta_n \sim \mathrm{Exp}(L\bar{\rho}_0(x_n))$ are exponentially distributed random numbers. Note that this can be viewed as a discrete time random walk. As $L\to\infty$ the $\eta_n = \mathcal{O}(1/L)$ become very small, meaning one can ignore locally the $x$ dependence of $\bar{\rho}_0(x_n)$. This stochastic process then locally generates a Poisson point process with the average density $\bar{\rho}_0(x)$. The slow dependence of $\bar{\rho}_0(x)$ on $x$ then creates variations of the density on the macroscopic scale. For practical purposes the procedure has to be stopped eventually, which we do once the density $\bar{\rho}_0(x) < 0.01$ is below a threshold.

The above procedure generates the particle locations for $x>0$. In order to generate the negative particle locations $x<0$ as well, we simply repeat the algorithm, but run it in the negative direction.

\subsection{Time evolution of the particles}
The particles can then be time-evolved to any time $t$ by solving Eq. \eqrefmt{2} for given $y_i(t) = y_i-\theta_i t$ numerically (the initial $y_i$ can be computed from Eq. \eqrefmt{2}). Numerically solving non-linear equations is a standard, but still non-trivial task. Appropriate methods could vary depending on the model, but here we choose to use the gradient descent algorithm. Its validity is guaranteed for models with positive phase-shifts, like the Lieb-Liniger phase shift -- more precisely, whenever $\p_x \p_\theta\psi(x,\theta\geq0$. It is based on the observation that Eq.~\eqrefmt{2} is the minimization condition of the following convex function:
\begin{align}
    S(x_1,\ldots, x_n) = \frac{1}{2}\sum_i (x_i-y_i)^2 + \frac{1}{2}\sum_{i\neq j} \gamma(x_i-x_j,\theta_i-\theta_j), 
\end{align}
where $\partial_x\gamma(x,\theta) = \partial_\theta\psi(x,\theta)$. In our case $\gamma(x,\theta) = \sqrt{x^2+\alpha^2}\varphi(\theta)$.

Since $S(x_1,\ldots, x_n)$ is a convex function the gradient descent algorithm for finding its minimizer is guaranteed to converge.  

\subsection{GHD simulations}
The GHD simulations were done using the IFluid package~\cite{10.21468/SciPostPhys.8.3.041}, which already implements the Lieb-Liniger model. The Euler scale GHD equation is solved using a Backward Semi-Lagrangian Implicit Runge-Kutta 4 method (this is a fourth order method, for details see~\cite{MOLLER2023112431}). The simulations were done on a $\mathrm{space}\times \mathrm{momentum} = 200 \times 200$ grid spanning $x \in [-8,8]$ and $\theta \in [-2,2]$. Time is discretized in units $\Delta t = 0.0025$.

\section{Change of metric and physical momentum distribution}

Here we describe a different method to derive the GHD equation which is based on the change of coordinates Eq.~\eqrefmt{2}. We can rewrite Eq.~\eqrefmt{2} directly in terms of the empirical density as (omitting the time argument)
\beq
	\b y_i
	= \b x_i + \int\dd{\b x}\dd \theta\, \rho_{\rm e}(\theta,\b x) \partial_\theta\psi(L(\b x_i-\b x),\theta_i-\theta)
\eeq
where in particular we use the fact that $\psi_\theta(0,0)=0$ by anti-symmetry. Assuming that $\b x_i = \b x(\theta_i,\b y_i)$ for some well-behaved function $\b x(\theta,\b y)$ and taking the large-$L$ limit, 
\beq
	 \b y =
	 \b x(\theta,\b y) + \int\dd{\b x}'\dd \theta'\, \rho_{\rm p}(\theta',\b x') \psi^{\sgn(\b x(\theta,\b y)-\b x')}\varphi(\theta-\theta').\label{equ:sm_GHD_y}
\eeq
The differential at fixed $\theta$ is
\beq\label{yx}
	\dd \b y =  (1 +  \int\dd \theta'\, \rho_{\rm p}(\theta',\b x) \varphi(\theta-\theta'))\dd \b x
	= 2\pi \rho_{\rm s}(\theta,\b x)\dd \b x
\eeq
where $\rho_{\rm s}(\theta,\b x)$ is the spatial, or total, density (with the conventional normalisation used in quantum systems). Eq.~\eqref{yx} is the transformation of coordinates $\b x\mapsto \b y$ that is known to trivialise the GHD equation \cite{Doyon2018}. This trivialised GHD equation can be solved immediately  and its solution can be mapped to the actual particle density by inverting Eq. \eqref{equ:sm_GHD_y}. Thus, as the $\b y_i$ coordinates indeed evolve trivially, this can form a basis for an alternative way of proving the emergence of the GHD equation.

In fact, a similar calculation may be made to obtain the physical momentum density. Using
\beq
	L\p_x \psi(x,\theta)\Big|_{x=L\b x} \to
	\delta(x)\phi(\theta)
\eeq
where $\phi(\theta) = \int_0^\theta\dd\theta'\,\varphi(\theta')$, we get from \eqrefmt{3}
\beqa
	p_i &=& \theta_i + L\int \dd \b x\dd\theta\rho_{\rm e}(\theta,\b x) \p_x\psi(L(\b x_i-\b x),\theta_i-\theta)\n
	&\to& \theta_i + \int\dd\theta\,
	\rho_{\rm p}(\theta,\b x_i)\phi(\theta_i-\theta)
\eeqa
from which we deduce $p_i = p^{\rm Dr}(\theta_i,\b x_i)$ where $p^{\rm Dr}(\theta,\b x)$ is the Dressed momentum (see e.g.~\cite{PhysRevLett.124.140603}); the latter is the physical momentum variation upon adding a particle of rapidity $\theta$ in Bethe ansatz systems (here in the state at macroscopic position $\b x$). Therefore, changing variable, the physical momentum distribution of our classical particle system is
\beq
	\rho_{\rm phys}(p,\b x) = L^{-1}\sum_i \delta(\b x-\b x_i)\delta(p-p_i)
	= \frc{\rho_{\rm p}(\theta(p,\b x),\b x)}{2\pi\rho_{\rm s}(\theta(p,\b x),\b x)}
\eeq
where we used $2\pi\rho_{\rm s}(\theta,\b x) = \p p^{\rm Dr}(\theta,\b x)/\p\theta$, and $\theta(\cdot,\b x)$ is the inverse of $p^{\rm Dr}(\cdot,\b x)$, i.e.~$p^{\rm Dr}(\theta(p,\b x),\b x) = p$. Using the expression for the occupation function $n(\theta,\b x) = \rho_{\rm p}(\theta,\b x)/\rho_{\rm s}(\theta,\b x)$, this reproduces the formula quoted in the main text.

\section{Trajectories from \ttbar-deformations}

In \cite{doyon2023generalised} we prove that the Hamiltonian Eq.~\eqrefmt{4} is well defined and arises from a generalised \ttbar-deformation of the system of free particles as per Eq.~\eqrefmt{6}, under certain conditions on $\psi$, including $\p_x\p_\theta\psi(x,\theta)\geq0$. There, we show that these conditions are sufficient to guarantee that Eqs.~\eqrefmt{2}, \eqrefmt{3} have unique solutions for $\bm x$ and $\bm \theta$, respectively. If the conditions is broken, there is in fact no guarantee that Eq.~\eqrefmt{3} can be inverted to define the Hamiltonian as a function of $\bm x,\bm p$. Likewise, there is no guarantee that Eq.~\eqrefmt{2} can be inverted to provide the trajectories. Nevertheless, we discussed in the main text how to make sense of trajectories in cases where invertibility does not hold: Eq.~\eqrefmt{2} still defines the trajectories, but one must interpret it correctly; one interpretation is that one must modify phase space in order to account for different particle numbers, and particle creation / annihilation occurs.

Here, we show that, indeed, \ttbar-deformations of free-particle trajectories give Eq.~\eqrefmt{2} without further conditions on $\psi(x,\theta)$: {\em any solution to Eq.~\eqrefmt{2} is compatible with the corresponding generalised \ttbar-deformation, even when Eq.~\eqrefmt{2} is not invertible}. Thus, the interpretation above is sensible; in particular, one may augment the phase space, and at critical values of $\lambda$, where multiple solutions arise for some times, it is consistent to consider these multiple solutions at once, interpreted as the creation of particles and anti-particles. A similar construction from Eq.~\eqrefmt{3}, to construct the full Hamiltonian on this larger phase space, is left for future works.

We note that the generalised \ttbar-deformation Eq.~\eqrefmt{5} can be obtained from the generator \cite{Pozsgay2020,10.21468/SciPostPhys.9.5.078,cardy2022t,doyon2023generalised}
\beq
	X^\lambda(\bm x,\bm p) = -\sum_{ij} \int_{x_i}^\infty \dd z\,w(z-x_j,\theta_i^\lambda(\bm x,\bm p)-\theta_j^\lambda(\bm x,\bm p)).
	\label{X}
\eeq
This is in the sense that the Hamiltonian is deformed as $H_\lambda \to H_\lambda + \delta\lambda\{H_\lambda,X^\lambda\}$. In fact, any conserved quantity transforms in this way, including the asymptotic momenta $\theta_i^\lambda(\bm x,\bm p)$.

As the flow in $\lambda$ is a Poisson flow, it preserves the Poisson algebra, thus it is natural to define the asymptotic impact parameters as those obtained from his flow:
\beq\label{yflow}
	y_i^\lambda(\bm x,\bm p) \to y_i^\lambda(\bm x,\bm p) + \delta\lambda \,\{y_i^\lambda(\bm x,\bm p),X^\lambda(\bm x,\bm p)\}.
\eeq

We now show that: {\em if Eq.~\eqrefmt{2} holds, in which we replace $\psi\to \psi^\lambda=\lambda\psi$, as well as $y_i,\theta_i\to y_i^\lambda,\theta_i^\lambda$, and if $\theta_i^\lambda$'s satisfy the flow equation, then the impact parameters $y_i^\lambda$'s satisfy the flow equation \eqref{yflow}, where the flow is defined using $w = \p_x\psi / 2$} (this is in agreement with Eq.~\eqrefmt{6}). Therefore, any solution to Eq.~\eqrefmt{2} is compatible with the corresponding generalised \ttbar-deformation.

Note that we do not need to define the Hamiltonian for the above statement to make sense: we directly use the flow on trajectories, much as was done in \cite{cardy2022t}. Note also that under the above identification, we have the asymptotic condition $|x|w(x,\theta)\to0$ ($|x|\to\infty)$, thus the function $X^\lambda(\bm x,\bm p)$ is well defined. In fact it takes the form
\beq
	X = \frc12\sum_{ij} \psi(x_{ij},\theta_{ij})
\eeq
where here and below we omit the superscript $\lambda$ and use the condensed notation $x_{ij} = x_i-x_j$, etc., for lightness of notation.

The proof is as follows. Under the flow, the left-hand side of Eq.~\eqrefmt{2} gives (we use indices $x,\theta$ to denote derivatives)
\beq
	\{y_i, X\} = 
	\sum_{jk}\{y_i,x_j\}\psi_{x}(x_{jk},\theta_{jk})+
	\sum_{j}
	\psi_{\theta}(x_{ij},\theta_{ij}).
\eeq
We may use Eq.~\eqrefmt{2} to obtain
\beq
	\{y_i,x_j\} = 
	\sum_l \psi_{\theta\theta}(x_{il},\theta_{il})
	\{\theta_{il},x_j\}
\eeq
whence
\beq\label{p1}
	\{y_i,X\}
	=
	\sum_{jkl} \psi_{\theta\theta}(x_{il},\theta_{il})
	\psi_x(x_{jk},\theta_{jk})\{\theta_{il},x_j\}+
	\sum_{j}
	\psi_{\theta}(x_{ij},\theta_{ij}).
\eeq
On the other hand, differentiating the right-hand side of Eq.~\eqrefmt{2}, with $\psi^\lambda = \lambda\psi$, we obtain
\beq\label{p2}
	\frc{\p y_i}{\p\lambda}
	= \sum_j \psi_\theta(x_{ij},\theta_{ij})
	+ \sum_l \psi_{\theta\theta}(x_{il},\theta_{il})
	\frc{\p \theta_{il}}{\p\lambda}.
\eeq
We evaluate the derivatives using the flow $\p\theta_i/\p\lambda = \{\theta_i,X\}$, to obtain
\beq\label{p3}
	\frc{\p \theta_{il}}{\p\lambda}
	=
	\sum_{jk}\{\theta_{il},x_j\}\psi_x(x_{jk},\theta_{jk}).
\eeq
Combining \eqref{p1}, \eqref{p2} and \eqref{p3}, we find
\beq
	\{y_i,X\} = \frc{\p y_i}{\p\lambda}
\eeq
which shows the statement.

\bibliographystyle{apsrev4-1}
\bibliography{bib.bib}


